\documentclass[11pt,twoside]{article}
\usepackage{iap2000,epsfig}

\def\prl{\em Phys. Rev. Lett.}

\def\aap{\em Astron. Astrophys.}

\def\be{\begin{equation}}
\def\ee{\end{equation}}
\def\bea{\begin{eqnarray}}
\def\eea{\end{eqnarray}}

\def\sze{SZE}
\def\apjl{\em ApJ}
\def\apj{\em ApJ}
\def\araa{\em Ann. Rev. Astron. Astrophys.}
\def\mnras{\em M.N.R.A.S.}
\def\aj{\em A.J.}
\def\nat{\em Nature}

\newcommand{\Ho}{\mbox{$H_0$}}
\newcommand{\neo}{\mbox{$n_{e  0}$}}
\newcommand{\dTo}{\mbox{$\Delta T_0$}}
\newcommand{\Sxo}{\mbox{$S_{x  0}$}}
\newcommand{\Sx}{\mbox{$S_{x  0}$}}

\newcommand{\LameH}{\mbox{$\Lambda_{e \mbox{\tiny H}}$}}

\newcommand{\Da}{\mbox{$D_{\!\mbox{\tiny A}}$}}
\newcommand{\kms}{\mbox{km s$^{-1}$}}
\newcommand{\ksM}{\mbox{km s$^{-1}$ Mpc$^{-1}$}}
\newcommand{\OmM}{\mbox{$\Omega_M$}}
\newcommand{\OmL}{\mbox{$\Omega_\Lambda$}}
\def\lsim{\mathrel{\hbox{\rlap{\hbox{\lower4pt\hbox{$\sim$}}}\hbox{$<$}}}}
\def\gsim{\mathrel{\hbox{\rlap{\hbox{\lower4pt\hbox{$\sim$}}}\hbox{$>$}}}}

\begin{document}

\title{The Sunyaev-Zel'dovich Effect: Results and Future Prospects}

\author{J.~E.~Carlstrom~\footnote{Dept.\ of Astronomy and Astrophysics, University of Chicago, 5640 S.\ Ellis Ave., Chicago, IL  60637}, 
        M.~Joy \footnote{Space Science Laboratory, SD50, NASA Marshall Space Flight Center,
Huntsville, AL 35812},  
	L.~Grego \footnote{Harvard-Smithsonian Center for Astrophysics, 60 Garden Street,
Cambridge, MA 02138}, 
        G.~Holder $^a$,\\ 
        W.~L.~Holzapfel \footnote{Dept.\ of Physics, University of California, Berkeley,
Berkeley, CA  94720}, 
        S.~LaRoque $^a$, 
        J.~J.~Mohr \footnote{Depts.\ of Astronomy, Physics, University of Illinois,
Urbana, IL 61801} and  
        E.~D.~Reese $^a$\\ }

\address{to appear in ``Constructing the Universe with Clusters of Galaxies",\\
IAP conference, July 2000, eds.~F.~Durret and G.~Gerbal, \\ 
http://www.iap.fr/Conferences/Colloque/coll2000/}

\maketitle\abstracts{ The Sunyaev-Zel'dovich effect (SZE) provides a
powerful tool for cosmological studies. Through recent advances in
instrumentation and observational techniques it is now possible to
obtain high quality measurements of the effect toward galaxy
clusters. The analysis of the SZE toward a few tens of clusters has
already led to interesting constraints on the Hubble constant and the
mass density of the universe. In the near future, instruments
exploiting the redshift independence of the SZE will be used to
conduct deep surveys for galaxy clusters providing detailed
information on the growth of large scale structure, tests of
cosmological models and tight constraints on the cosmological
parameters that describe our universe. In this review we provide an
overview of the SZE and its use for cosmological studies. We summarize
the current state of observations and the constraints on cosmological
parameters already obtained and we discuss the power of using the
SZE for future deep cluster surveys.}

\section{Introduction}

\label{sec:intro}

The Sunyaev-Zel'dovich Effect (SZE) provides a unique and powerful
observational tool for cosmology. It is particularly useful for
determining cosmological parameters when combined with other
observational diagnostics of clusters of galaxies such as X-ray
emission from the intracluster gas, weak and strong lensing by the
cluster potential, and optical galaxy velocity dispersion
measurements.  For example, cluster distances can be determined from
the analysis of SZE and X-ray data, leading to an estimate of the Hubble
constant and eventually the deceleration parameter. 
The baryonic and total masses of clusters can be measured directly, allowing
an estimate of the ratio of the baryonic to total matter density of the universe.

All of the measurements provide insights into the structure of galaxy
clusters themselves. Furthermore, many of the cluster properties
derived directly from a given observation, or from a combination of
observations, can be determined in several different ways. For
example, the gas mass fraction can be determined by various
combinations of SZE, X-ray, and lensing observations. The electron
temperature, a direct measure of a cluster's mass, can be measured
directly through X-ray spectroscopy, or determined through the
analysis of various combinations of X-ray, SZE, and lensing
observations.  Many of the desired properties of clusters are
therefore over-constrained by observation, allowing a deeper
understanding of clusters and critical tests of current models
for the formation and evolution of galaxy clusters.

Perhaps the most unique and powerful cosmological tool provided by the
exploitation of the SZE is the possibility to conduct deep large scale surveys for galaxy
clusters. SZE observations are particularly well suited for deep
surveys because the detection threshold for such a survey depends on 
the mass of
the cluster and is independent of distance. {\it SZE surveys will be
able to detect all clusters above a mass limit independent of the
redshift of the clusters.}  This remarkable property of SZE surveys is
due to the fact that the SZE is a distortion of the CMB
spectrum. While the CMB suffers cosmological dimming with redshift,
the ratio of the SZE to the CMB does not; it is a
direct, redshift independent measurement of the intracluster medium
(ICM) column density weighted by temperature, i.e., the pressure
integrated along the line of sight. The total SZE flux detected will
be proportional to the total temperature-weighted mass (total
integrated pressure) and inversely proportional to the
square of the angular diameter distance. Adopting a reasonable
cosmology and accounting for the increase in the universal matter
density with redshift, the mass limit for a given SZE survey flux
sensitivity will rise with redshift, level off at redshifts
approaching one and actually decrease at redshifts greater than one.

There has been considerable progress recently in detecting and imaging
the SZE.  Efforts over the first two decades after the SZE was first
proposed in 1970 \cite{sunyaev70,sunyaev72} yielded only a couple of
reliable detections.  Over the last decade new detectors and observing
techniques have allowed high quality detections and images of the
effect for more than 40 clusters with redshifts as high as 0.9. Now at
the start of the fourth decade we are in position to exploit fully the
power of the SZE by obtaining detailed images of a set of clusters to
understand the ICM, by observing large SZE samples of clusters to
determine statistically robust estimates of the cosmological
parameters, and to commence large untargeted SZE
surveys to probe the high redshift universe. These surveys will
provide a direct view of the growth of large scale structure and allow
constraints to be placed on the equation of state of the dark
energy.\cite{holder00,haiman00}

In this brief review, we first outline the properties of the SZE and
its use for cosmological studies. Next we provide an overview of the
current observational techniques with examples of recent results and a
summary of observations obtained to date. The constraints that have
been placed on cosmological parameters are given next. Finally,
we discuss the expectations for future SZE observations and their
role in cosmological studies.

\section{The Sunyaev-Zel'dovich Effect}

\begin{figure}[!t]
\epsfxsize= 4.5 in
\centerline{\epsfbox{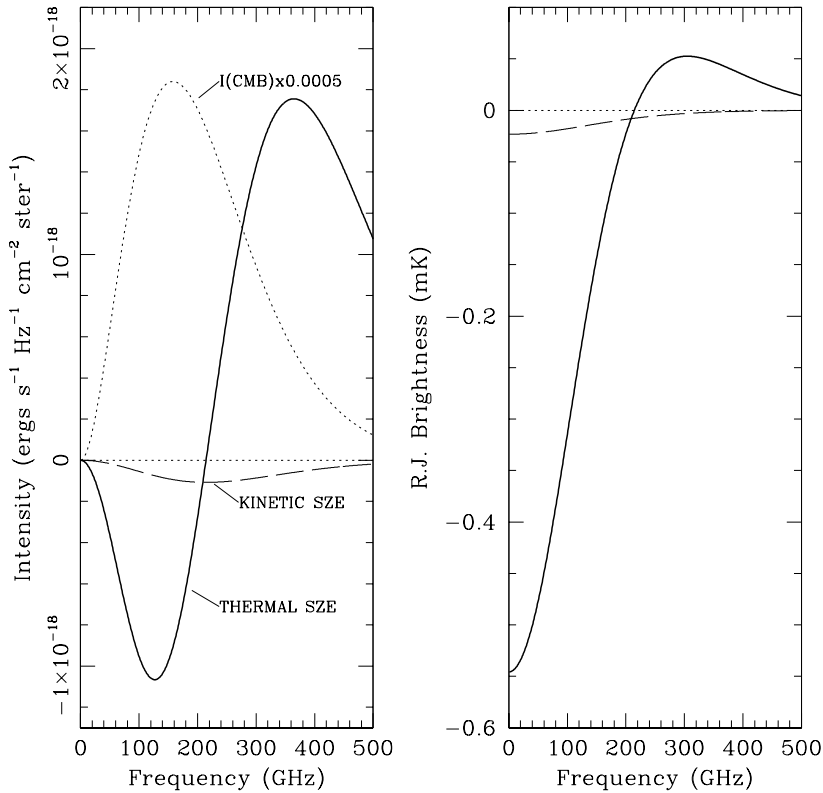}}
\caption{Spectral distortion of the cosmic microwave background (CMB)
radiation due to the Sunyaev-Zel'dovich effect (SZE). The left panel 
shows the intensity and the right panel shows the Rayleigh-Jeans 
brightness temperature. The thick solid line is the thermal SZE and
the dashed line is the kinetic SZE. For reference
the 2.728~K Planck spectrum of the CMB \cite{fixsen96} (scaled by 0.0005) is 
shown by the dotted line
in the left panel; a line at zero is shown in both panels. The 
cluster properties used to calculate the 
spectra are an electron temperature of 10~keV, a Compton
$y$~parameter of $10^{-4}$ and a peculiar velocity of 500 \kms.}
\label{fig:spectrum}
\end{figure}

The Sunyaev-Zel'dovich effect (SZE) is a small spectral distortion of
the cosmic microwave background (CMB) caused by the
scattering of CMB photons off a distribution of high energy
electrons. Here we concentrate on the SZE caused by the hot thermal
distribution of electrons provided by the intracluster medium (ICM) of
massive galaxy clusters.  The average energy exchange in the
scattering of a CMB photon off an electron is $k_BT_e / m_e c^2$.

The derivation of the \sze\ can be found in the original papers of
Sunyaev and Zel'dovich,\cite{sunyaev70,sunyaev72} in several
reviews,\cite{sunyaev80,rephaeli95,birkinshaw99} and in a number of
more recent contributions which include relativistic 
corrections.\cite{rephaeli95,rephaeli97,itoh98,challinor98,nozawa98b,sazonov98}

The SZE distortion of the CMB spectrum 
at dimensionless frequency
$x \equiv \frac{h\nu}{k_BT_{CMB}}$ is given by
\begin{equation}
   \frac{\Delta T_{SZE}}{T_{CMB}} =
   f(x,T_e) \ y  ,
   \label{eq:deltaT1}
\end{equation}
where $y$ is the Compton $y$-parameter equal to the optical depth,
$\tau_e = \sigma_T \int n_e dl$,
times the fractional energy gain per scattering
\begin{equation}
  y = \sigma_T \int n_e \frac{k_B T_e}{m_e c^2} d\ell, 
  \label{eq:y}
\end{equation}
with $\sigma_T$ the Thomson cross-section, $n_e$ the electron number
density, $T_e$ the electron temperature, $k_B$ the Boltzmann constant,
$m_e c^2$ the electron rest mass energy and the integration is along
the line of sight.

The spectral distortion as a function of the dimensionless frequency
is given by
\begin{equation}
   f(x,T_e) = \left(x \frac{e^x+1}{e^x-1} -4\right)\left(1 + \delta_{SZE}(x,T_e)\right),
   \label{eq:fx}
\end{equation}
where $\delta_{SZE}(x,T_e)$ is the relativistic correction to the
frequency dependence.  Note that $f(x,T_e) \rightarrow -2$ in the
non-relativistic and Rayleigh-Jeans (RJ) limits.

If the cluster is moving with respect to the CMB rest frame with a
component of its velocity, $v_{pec}$, projected along the line of sight to the
cluster, then the observed distortion of the CMB spectrum will have a component due to
the Doppler effect referred to as the kinetic SZE. In
the non-relativistic limit, the spectral signature of the kinetic SZE
is a purely thermal distortion of magnitude 
\begin{equation}
\Delta T/T_{CMB}  = -\tau_e (v_{pec}/c),
\label{eq:v_pec}
\end{equation}
where $v_{pec}$ is along the line of sight; the emergent
spectrum is still described completely by a Planck spectrum, but at a
slightly different temperature, lower (higher) for positive (negative)
peculiar velocities.\cite{sunyaev72,phillips95,birkinshaw99}

Fig.~\ref{fig:spectrum} illustrates the spectral distortions of the
CMB induced by the thermal and kinetic SZE of a massive cluster.  The
left panel shows the change in intensity and the right panel shows the
change in Rayleigh-Jeans (RJ) brightness temperature. The RJ
brightness is shown because the sensitivity of a radio telescope is
calibrated in these units. It is defined simply by $I_\nu = (2 k_B
\nu^2/c^2) T_{RJ}$ where $I_\nu$ is the intensity at frequency $\nu$,
$k_B$ is Boltzmann's constant and $c$ is the speed of light. The
Planck spectrum of the CMB radiation is also shown by the dotted line
(multiplied by 0.0005) for reference. Note that the thermal effect
causes a decrement at low frequencies ($\nu < 218$~GHz) and an
increment at high frequencies and is distinguished readily from a
simple temperature fluctuation of the CMB.  The kinetic SZE distortion
is shown by the dashed curve. In the non-relativistic regime, it is
indistinguishable from a CMB temperature fluctuation.

The gas temperatures measured in massive galaxy clusters are around
$k_B T_e \sim 10$ keV.\cite{mushotzky97,allen98} At these
temperatures, electron velocities are becoming relativistic and small
corrections are required to calculate the proper SZE. For a massive
cluster with $k_BT_e/m_ec^2 \sim 0.02$ the relativistic corrections to
the SZE are of order a few percent in the RJ portion of the spectrum,
but can be substantial near the null of the thermal effect. Convenient
analytical approximations to fifth order in $k_BT_e/m_e c^2$ are
presented in Itoh et al.\cite{itoh98}

Relativistic perturbations to the kinetic SZE are due to the Lorentz
boost to the electrons provided by the bulk
velocity.\cite{nozawa98b,sazonov98} The leading term is of order $(k_B
T_e/m_e c^2)(v/c)$ and for a 10~keV cluster moving at 1000 \kms\ the
effect is about an 8\% correction. The $(k_B T_e/m_e c^2)^2(v/c)$ term
is about 1\%, and the $(v/c)^2$ term is only 0.2\%.

\section{Cosmology with the Sunyaev-Zel'dovich Effect}
\label{sec:cosmo}

The unique aspect of using the SZE for cosmology is that it is simply
a measure of the column density of electrons weighted by
temperature. 
The total SZE flux of a
cluster is also a clean measure of the total energy of the cluster.

The SZE is perhaps best known for enabling an independent
determination of the Hubble constant as described below.  More
recently, SZE data has been used to determine cluster gas mass
fractions which in turn have led to constraints on the universal
matter density parameter $\Omega_M$. While these applications are of
substantial interest, the availability of high quality SZE data offers
the ability to learn a great deal more about clusters and cosmology.

Armed with data sets from multiple observables of a clusters, many
parameters of interest, i.e., masses, gas mass fraction, electron temperature,
Hubble constant, gas clumping factor, are often over-constrained. This
will allow assumptions of the ICM structure to be tested, providing a
much better understanding of galaxy clusters. In turn, this will lead
to improvements in the precision with which cluster observations can
be used to determine cosmological parameters.

For example, it is assumed that thermal pressure is the only
significant form of support when the virial theorem is used to derive
cluster masses.  There is currently no compelling evidence against
this assumption.  A comparison of masses derived from the virial
theorem (using X-ray, SZE or galaxy velocity dispersion data) and
gravitational lensing should be able to test this
assumption.\cite{miralda-escude95b,loeb94,wu97,squires96,allen98}

As the precision with which we know cosmological parameters increases,
the same calculations now used to derive the parameters can be turned
around to solve for cluster properties. For example, the masses and
electron temperatures of distant high redshift clusters ($z > 0.8$)
were determined directly from SZE observations without access to X-ray
data,\cite{joy01} based on the assumption that their gas mass
fractions are consistent with the mean gas fraction derived from SZE
and X-ray spectroscopic data for a sample of 18 high redshift
clusters.\cite{grego99,grego00b} The ability to determine cluster
temperatures and masses from only SZE data without access to X-ray
data has important consequences for the analysis of deep SZE surveys
of the distant universe.

SZE surveys for distant clusters will provide a powerful new tool for
cosmological studies.  As discussed below, SZE surveys will provide a
direct view of the distant universe by finding all clusters regardless
of redshift; should clusters exist at redshifts much higher than
currently predicted they will be found by SZE surveys, but missed in
even the deepest X-ray observations planned. The surveys will provide
a large sample of clusters which can be used, for example, to refine
the determination of the Hubble constant.  A powerful use of
SZE surveys will be to place tight constraints on $\Omega_M$ and
$\Omega_\Lambda$, and for sufficiently large samples, on the equation of
state of the dark energy from the analysis of the evolution 
of the number density and 
mass distribution of the survey yields.

\subsection{Distance Determinations, Hubble constant}

\label{sec:cosmo-hubble}

Several years after the SZE was first proposed
\cite{sunyaev70,sunyaev72} it was recognized that the distance to a
cluster, and therefore the Hubble constant, could be determined with a measure of its SZE and its X-ray
emission.\cite{cavaliere77,gunn78,silk78,cavaliere78,boynton78,birkinshaw79}

The distance can be determined by exploiting the different density
dependencies of the SZE and X-ray emissivity.  The SZE is proportional
to first power of the density: $\Delta T \sim \int d\ell n_e T_e$,
where $d\ell$ is along the line of sight,  $n_e$ is the electron density and
$T_e$ is the electron temperature.  We define 
$d\ell \equiv \Da d\zeta$ to relate $\Da$ and an `angle' along the line of
sight to the angle subtended by the cluster on the sky. The X-ray emission is proportional
to the second power of the density: $\Sx \sim \int d\ell n_e^2
\LameH$, where \LameH\ is the X-ray cooling function.  The angular
diameter distance is solved for by eliminating the electron density
\footnote{Similarly, one could eliminate \Da\ in favor of the central density,
\neo} yielding
\begin{equation}
\Da \propto \frac{(\dTo)^2 \LameH}{\Sxo T_{e 0}^2} \frac{1}{\theta_c},
	\label{eq:Dadepend}
\end{equation}
where these quantities have been evaluated along the line of sight
through the center of the cluster (subscript 0) and $\theta_c$ refers
to a characteristic scale of the cluster along the line of sight,
whose exact meaning depends on the density model adopted.
  
We need to know (or assume) something
about the geometry of the cluster to relate an observation of 
$\theta_c^{sky}$ to the characteristic scale of the cluster along
the line of sight, $\theta_c$.  Typically, spherical symmetry is
assumed for the cluster geometry since for a large sample of clusters
one would expect $\left<\theta_c^{l.o.s.}/\theta_c^{sky}\right> = 1$,
at least in the absence of selection effects (e.g., \cite{sulkanen99}). 
A survey of a few hundred clusters with redshifts extending
beyond one would allow the technique to be used to trace the expansion
history of the universe, providing a valuable independent check of the
type Ia supernova results.\cite{riess98,perlmutter99} Such a survey is
now observationally well within reach.

It should be noted that in the SZE determination of the distance to a
cluster, it is assumed 
that the clumping factor $C \equiv
\left<n_e^2\right>^{1/2}/\left<n_e\right>$, equals unity. 
The derived Hubble constant will be a factor of
$C^2$ times larger than the true Hubble constant.

\subsection{Cluster Peculiar Velocities}

\label{sec:cosmo-vpec}

The line of sight velocity of a cluster with respect to the CMB rest
frame, the peculiar velocity of the cluster, can be measured by
separating the kinetic from the thermal \sze.  From inspection of
Fig.~\ref{fig:spectrum}, it is clear that this is best done by
observation at frequencies near the null of the thermal effect at
$\sim 218$~GHz.

Such measurements offer the ability to measure the peculiar velocity
of clusters at high redshifts which could be used to constrain
large scale gravitational perturbations to the Hubble flow.  The
intrinsic weakness of the effect makes it challenging to observe.
Upper limits have been placed on the peculiar velocities of 
clusters,\cite{holzapfel97b,laroque01} but a clear detection of the kinetic
effect has not yet been obtained. The kinetic \sze\ is a unique and
potentially powerful cosmological tool as it provides the only known
way to measure large scale velocity fields at high redshift.  As
discussed in section~\ref{sec:confusion}, contamination by CMB
temperature fluctuations as well as other sources will make it
difficult to determine accurately the peculiar velocity for a given
cluster.  It may be possible, however, to determine mean peculiar
velocities on extremely large scales by averaging over many clusters.

\subsection{Baryon mass fraction of clusters, $\Omega_M$, and $T_e$ }

\label{sec:cosmo_f_gas}

The total SZE flux from a cluster is simply proportional to the gas
mass of the cluster weighted by the temperature of the gas ($\int
d\Omega \Delta T_{SZE} \sim M_{gas} \left<T_e\right> \Da^{-2}$).  Combined with a measure
of the total gravitating mass of the cluster, one can estimate the
fraction of the mass contained in the ICM. The ICM contains most of
the baryons confined to the cluster potential with roughly an order of
magnitude more baryonic mass than that observed in the galaxies
themselves.\cite{white93,forman82} The gas mass fraction is
therefore a reasonable estimate of the baryonic mass fraction of the
cluster. Furthermore, it is not believed that segregation
between baryonic and non-baryonic mass occurs
on the scales from which massive clusters condense ${\rm \sim 1000\
Mpc^3}$, although a small fraction of baryons ($\sim$15\%) are likely
lost during the cluster formation process.\cite{white93} The baryonic
mass fraction of clusters therefore should reflect the universal mass
fraction of baryons to total matter, $\Omega_B/\Omega_M$ where $\Omega
\equiv \rho/\rho_c$ and $\rho_c$ is the critical density of the
universe.

With knowledge of $\Omega_B$, the measured gas mass fraction of a
cluster of galaxies leads to an estimate of the universal matter
density, $\Omega_M$. Recent reanalysis of BBN predictions with careful
uncertainty propagation \cite{nollett00,burles99} along with recent D/H
measurements in Ly$\alpha$ clouds \cite{burles98a,burles98b} constrain
the baryon density to be $\Omega_B h_{100}^2 = 0.019 \pm 0.0012$ at 68\%
confidence.  An additional independent determination of $\Omega_B h^2$
will soon be provided by precision measurements of the intrinsic CMB
anisotropy angular power spectrum.

The gas mass is measured directly by observations of the SZE provided
the electron temperature is known. The total gravitating mass can be
determined through hydrostatic equilibrium 
with knowledge of the
distribution of the gas and, again, knowledge of the electron
temperature. The SZE derived gas mass fraction will therefore be
proportional to $\Delta T_{SZE} / T_e^2$. Alternatively, the total
gravitating mass can be determined from measurement of strong lensing (on small scales)
or from weak lensing (on large scales).

A comparison of SZE derived gas mass and lensing total gravitating
mass is particularly interesting as both are measures of the projected
mass distribution. It has been shown that the gas mass fraction can be
determined from the analysis of SZE and lensing measurements without need to
parameterize the ICM distribution.\cite{holder00c} Furthermore, 
by comparing this mass
fraction with one derived using the virial theorem, it is possible to
solve for the ICM electron temperature.\cite{holder00c}

Cluster gas mass fractions can also be determined from
cluster X-ray emission in the same manner as from SZE measurements.
There are several important differences, however.  First the X-ray
emission, being proportional to the ICM density squared, is more
susceptible to clumping of the gas, $C$. 
The X-ray derived gas mass is essentially insensitive to the electron
temperature, while the SZE derived gas mass is proportional to $1/T_e$.

Currently, X-ray data for low redshift clusters is of exceptional
quality, far surpassing SZE data.  X-ray based gas mass fractions have
been measured to cluster radii of 1 Mpc or 
more.\cite{white95,david95,neumann97,squires97,mohr99} 
In an X-ray
flux-limited sample of 45 clusters,\cite{mohr99} the mean
cluster gas mass fraction within approximately the virial radius 
was found to be
(0.0749 $\pm 0.0005) h_{100}^{-3/2}$.  The quality of SZE images is
improving quickly as discussed in section~\ref{sec:observations}, and
already the SZE and X-ray data for high redshift clusters are of
comparable quality (see Sec.~\ref{sec:observations}).

\subsection{Cosmology with SZE Surveys}

\label{sec:cosmo-survey}

Perhaps the most powerful use of the SZE for cosmology will be to
probe the high redshift universe.  Sensitive, non-targeted surveys of
large regions of the sky for the SZE will provide an accurate 
inventory of clusters independent of their redshift.

\begin{figure}[tbh]
\epsfxsize= 6 in
\centerline{\epsfbox{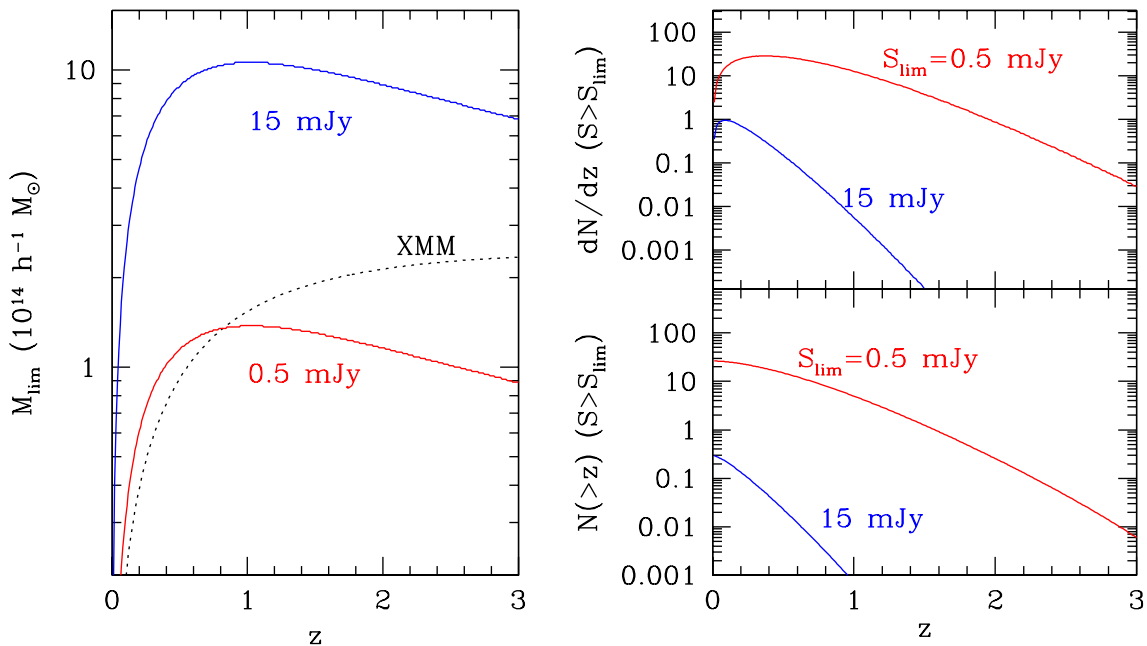}}
\caption{
Left: Mass limits as a function of redshift for a typical wide-field type 
of survey (sensitivity limit of $\sim$ 15 mJy at 30 GHz) and for a typical
deep survey ($\sim$ 0.5 mJy). The approximate XMM serendipitous
survey limit is also shown. Right: Differential (top) and cumulative
(bottom) counts per square degree as a function of redshift for two SZE surveys shown at
left, assuming a $\Lambda$CDM cosmology.\cite{holder00}
\label{fig:SZE_survey}}
\end{figure}

The evolution of the abundance of galaxy clusters is a sensitive probe
of cosmology. Measurements of the clusters masses and number density
as a function of redshift can be used to constrain the matter density,
$\Omega_M$,\cite{viana99,bahcall98,oukbir97,haiman00} and, for
sufficiently large samples, the equation of the state of the dark
energy.\cite{haiman00} X-ray surveys have already been used to
constrain $\Omega_M$, but they have been limited by sample size and
their reduced sensitivity to high redshift clusters.  SZE surveys
offer the attractive feature of probing the cluster abundance at high
redshift as easily as the local universe.

The sensitivity of a SZE survey is essentially a redshift independent
mass limit (see Fig.~\ref{fig:SZE_survey}).\cite{bartlett94,barbosa96,holder00}
The integrated SZE for a
cluster is proportional the temperature weighted mass of the cluster
and inversely to the square of the angular diameter distance, $\Delta
T_{SZE} \propto \left<T_e\right>M_{virial}/D_A^2$.  At low redshift
the mass detection threshold of a SZE survey increases with $D_A^2$.
At high redshift, $z \sim 1$, the mass threshold flattens and even
begins to decrease. This is due to two effects: 1) $D_A(z)$ begins to
flatten and 2) for a given virial mass, $M_{virial}$, high redshift
clusters will be denser and hotter (due to the increase in the matter
density) leading to a stronger SZE flux.

SZE surveys can be broadly grouped into two types of surveys: large
area, relatively shallow surveys, similar to what can be expected with
the Planck Surveyor Satellite, or relatively small, deep surveys, such
as those expected with upcoming ground-based efforts. In
Fig.~\ref{fig:SZE_survey} we show the approximate expected limiting
mass for the two types of surveys, assuming a simple limit on the
integrated flux density of 15 mJy at 30 GHz as expected for Planck and
0.5 mJy for deep ground-based surveys. We use 30 GHz as an arbitrary
normalizing frequency; to convert to integrated Compton $y$-parameter
use $S_{30} = 12.7 (y/arcm^2)$ Jy. As a comparison, we have also shown
an approximate mass limit for an XMM serendipitous 
survey,\cite{romer00} assuming a simple X-ray flux limit. A deep SZE survey
will be able to probe somewhat lower masses at redshifts past $z\sim
1$, which promises to be a very interesting regime.

In the right panels of the Fig.~\ref{fig:SZE_survey}, we show the
expected counts per square degree for a $\Lambda$CDM model
($\Omega_m=0.3, \Omega_\Lambda=0.7, h=0.65, \sigma _8=1$). A deep SZE
survey can expect to find roughly one cluster per square degree with
$z>1.5$.  On the other hand, even though a shallow survey has a
relatively high mass limit, a survey that covers half the sky would
find roughly 10000 clusters, many of which would be be near enough to
appear in the Sloan Digital Sky survey. 

\begin{figure}[!t]
\epsfysize= 3.0 in
\centerline{\epsfbox{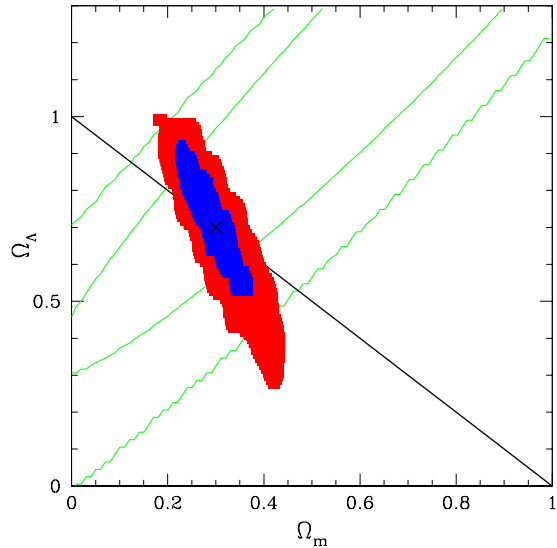}}
\caption{A simulation of the constraints in the $\Omega_M$--$\Omega_\Lambda$
plane which should be possible from the analysis of a 
SZE survey covering 12 square degrees in
which all clusters above $10^{14} h_{100}^{-1}$ M$_\odot$ are detected. 
The input cosmology is $\Omega_M=0.3$ and $\Omega_\Lambda =0.7$. The
blue region corresponds to the 68\% confidence region, red to
95\%. The value of $\sigma_8$ has been marginalized over rather than
kept fixed.  The diagonal ellipses are the current constraints based on the
analyses of type Ia supernovae at 68\% and 95\%
confidence.\cite{riess98,perlmutter99} The diagonal line at $\Omega_M
+ \Omega_\Lambda = 1$ is for a flat universe as suggested by recent
CMB anisotropy measurements.\cite{miller99,debernardis00,hanany00}}
\label{fig:om_ol}
\end{figure}

There are many possible uses of a cluster catalog extending to high
redshift with a simple selection function. An important use would be
to use the cluster abundance and its redshift evolution as a strong
test of cosmological parameters \cite{haiman00} and the structure
formation paradigm in general. In Fig.~\ref{fig:om_ol} we show the
expected constraints in the $\Omega_M$--$\Omega_\Lambda$ plane based on
the analysis of a hypothetical deep SZE survey \cite{carlstrom99,holder00} covering
12 square degrees in which all clusters above $10^{14} h_{100}^{-1}$
M$_\odot$ are detected and for which redshifts are known. 
The input cosmology is $\Omega_M=0.3$ and $\Omega_\Lambda =0.7$. The
blue region corresponds to the 68\% confidence region, red to
95\%. The value of $\sigma_8$ has been marginalized over rather than
kept fixed. Obtaining redshifts for the clusters found in such a survey 
would be greatly facilitated
by coordinating the selection of the SZE fields with those of deep
optical surveys. Redshifts would then be available
for all clusters with $z \lsim 1$. Deep follow-up optical and infrared observations
could then be targeted at the resulting list of high redshift clusters.

Another use of the resulting catalog is to probe the redshift
evolution of the intracluster medium, perhaps seeing the effects of
feedback from galaxy formation \cite{holder01} out to $z\sim 2$. In
fact, so little is currently known about relatively low-mass clusters
at high redshift that it is difficult to predict the most important
uses of such a catalog.

\section{Observations of the Sunyaev-Zel'dovich Effect}
\label{sec:observations}

In the twenty years following the first papers by Sunyaev and
Zel'dovich \cite{sunyaev70,sunyaev72} there were few firm detections
of the \sze\ despite a considerable amount of
effort.\cite{birkinshaw91} Over the last several years, however,
observations of the effect have progressed from low signal to noise 
detections and
upper limits to high confidence detections and detailed images. In
this section we briefly review the state of \sze\ observations. The
constraints on the cosmological parameters that have already been
determined by existing \sze\ observations are discussed in the next
section.

The dramatic increase in the quality of the observations is due to
improvements both in low-noise detection systems and in observing
techniques, usually using specialized instrumentation to carefully
control the systematics that often prevent one from obtaining the
required sensitivity.  Such systematics include, for example, the
spatial and temporal variations in the emission from the atmosphere
and the surrounding ground, as well as gain instabilities inherent to
the detector system used. To appreciate the importance of controlling
these systematics, consider the theoretical sensitivity of a modestly
sensitive radio receiver with a noise temperature \footnote{the
required input power expressed in temperature units that would result
in the same measured output noise power of an equivalent but noiseless
system} of 100~K and a detector bandwidth of 1~GHz; it would be $3 mK
s^{-1/2}$. Such a system should secure a detection of the SZE for a
massive cluster within an hour or so.  However, observations with far
more sensitivity have failed to detect the SZE even after many hours
of integration. Detector systems today typically have sensitivities
around $ 1 mK s^{-1/2}$. Clearly, the goal of all modern SZE
instruments is control of systematics.

The observations must be conducted on the appropriate angular scales.
Galaxy clusters are large objects with a characteristic scale size of
order a Mpc. In any reasonable cosmology, a Mpc subtends an
arcminute or more at any redshift; low redshift clusters will subtend
a much larger angle, for example the angular extent of the Coma
cluster ($z = 0.0235$) is of order a degree (core radius $\sim 10'$).
\cite{herbig95} The detection of extended low surface brightness
objects requires precise differential measurements made toward widely
separated directions on the sky. On these large angular scales,
offsets due to differential ground pick-up and
atmospheric variations are difficult to control.

\subsection{Sources of astronomical contamination and confusion}

\label{sec:confusion}

In designing an instrument for SZE observations, one needs to take
into account several sources of possible contamination and confusion
from astronomical sources. One such source is anisotropy of the CMB
itself. For distant clusters with angular extents of a few arcminutes
it is not a problem as the CMB anisotropy has been shown to be damped
considerably on these scales.\cite{holder99a,holzapfel00a,dawson01}
For nearby clusters, or for searches for distant clusters using large
beams of order 10$'$ or more, the intrinsic CMB anisotropy must be
considered. The unique spectral behavior of the thermal \sze\ can be
used to separate it from the intrinsic CMB in these cases. Note,
however, that for such cases it will not be possible to separate the
kinetic \sze\ effects from the intrinsic CMB anisotropy without
relying on the very small spectral distortions of the kinetic \sze\
due to relativistic effects.

Historically, the major source of contamination in the measurement of
the \sze\ has been radio point sources. It is obvious that emission
from point sources located along the line of the sight to the cluster
could fill in the \sze\ decrement, leading to an underestimate. The
radio point sources are variable and therefore must be
monitored. Radio emission from the cluster member galaxies, often from
the central CD galaxy, is often the largest source of radio point
source contamination, at least at high radio 
frequencies.\cite{cooray98}  The flux, $S_\nu$, of a
radio source typically has a spectral index of 
$\alpha \sim 0.7$ for $S_\nu = \nu^{-\alpha}$.  In the RJ limit, the \sze\ flux is
proportional to $\nu^2$ and therefore point sources are much less of an
issue at higher radio frequencies.

While it is most likely that insufficient attention to radio point
sources would lead to the underestimate of the \sze\ effect, it could
also lead to an overestimate. The most obvious example is if
unaccounted point sources are in the reference fields surrounding the
cluster.  An effect due to gravitational lensing has also been pointed
out for low frequency observations where the flux from many point
sources must be taken into account before a reliable measure of the \sze\ can
be made.  Essentially, the added efficiency of detecting
point sources toward the center of the cluster due to gravitational
lensing could lead to an overestimate of the \sze\
decrement.\cite{loeb1997} This effect should be negligible at
frequencies greater than roughly 30~GHz.

At frequencies near the null of the thermal \sze\ and higher, dust
emission from extragalactic sources as well as dust emission from our
own galaxy must be considered. Over the
frequencies of interest, dust emission rises steeply
as $\nu^{2+\beta}$, where $\nu^\beta$ is the scaling of the 
dust opacity with $1 < \beta < 2$. Consider the dusty extragalactic sources
that have been found toward massive galaxy clusters with the SCUBA
bolometer array.\cite{smail97} Sources with 350~GHz (850~$\mu m$)
fluxes greater than 8~mJy are common and all clusters surveyed had
multiple sources with fluxes greater than 5~mJy. A 10~mJy source at
350~GHz corresponds to a contaminant of the CMB of magnitude 
$\Delta T_{CMB} = 345\ \mu$K for 1$'$ beam, or
a Compton $y$-parameter of $6\times 10^{-5}$.  The same source scaled
to 270~GHz, assuming a dust opacity law of $\nu^{1.5}$, leads
to 
$\Delta T_{CMB} = 95\ \mu$K for a 1$'$ beam and a $y$-parameter of $4\times
10^{-5}$. Scaling to the SZE thermal null at 218 GHz gives 5 mJy which
corresponds to a $\Delta T_{CMB} = 41\ \mu$K for a 1$'$ beam. This in
turn translates directly to an uncertainty in a measurement of the
cluster peculiar velocity (Eq.~\ref{eq:v_pec}); for a massive cluster
with an optical depth of 0.01 ($y$-parameter of $2\times 10^{-4}$ and
an electron temperature of 10~keV), 41~$\mu$K corresponds to a peculiar
velocity of 450 \kms.  The contamination is more severe for less
massive clusters with the dependence scaling as $\Delta v_{pec}
\propto \tau_e^{-1} \propto R^2/M \propto M^{-1/3} \propto
T_e^{-1/2}$. 

As with SZE observations in the radio, high frequency observations
also need to consider the effects of point sources and require
either high dynamic angular range, large spectral coverage, or both,
to separate the point source emission from the SZE.

\subsection{Single Dish Observations}

\label{sec:single_dish}

\begin{figure}[tbh]
\epsfysize= 2 in
\centerline{\epsfbox{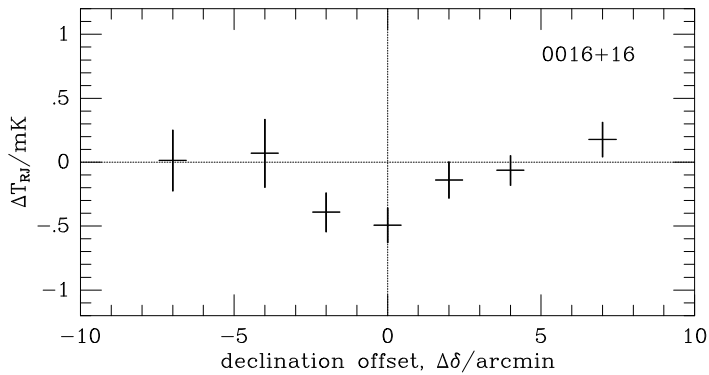}}
\caption{A measurement of the SZE profile
across the cluster CL~0016+16 obtained
with the OVRO 40 m telescope.\cite{birkinshaw91,hughes98}.
The observed profile
provided confidence in the reliability
of this early detection.}
\label{fig:birkinshaw}
\end{figure}

The first measurements of the \sze\ were made with single dish radio
telescopes at cm wavelengths.  Advances in detector technology made
these measurements possible, although early observations appear to
have been plagued by systematic errors which led to irreproducible and
inconsistent results.  Eventually, successful detections were
obtained, although the reported results show considerable scatter
reflecting the difficulty of these measurements.  During this period,
the pioneering work of Birkinshaw and collaborators with the OVRO 40
meter telescope stands out for its production of results which served
to build confidence in the
technique.\cite{birkinshaw78a,birkinshaw78b} One of the first
measurements of the profile of the SZE decrement across a galaxy
cluster is shown in
Fig.~\ref{fig:birkinshaw}.\cite{hughes98,birkinshaw86,birkinshaw91}

%
%

\begin{figure}[!t]
\epsfysize= 2.0 in
\centerline{\epsfbox{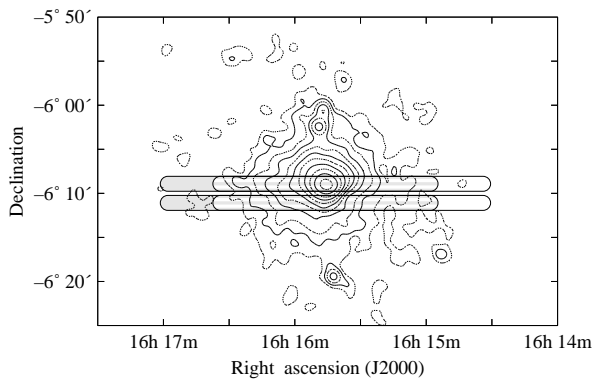}}
\caption{
The drift scans of the two detector rows of the SuZIE instrument
superimposed on the X-ray surface brightness contours for Abell~2163. The two
sets of scans, white and grey, correspond to two sets of measurements that
differed only in the position at which the drift scan 
started.\cite{holzapfel97a}}
\label{fig:SuZIEscans}
\end{figure}

\begin{figure}[!h]
\epsfysize= 4 in
\centerline{\epsfbox{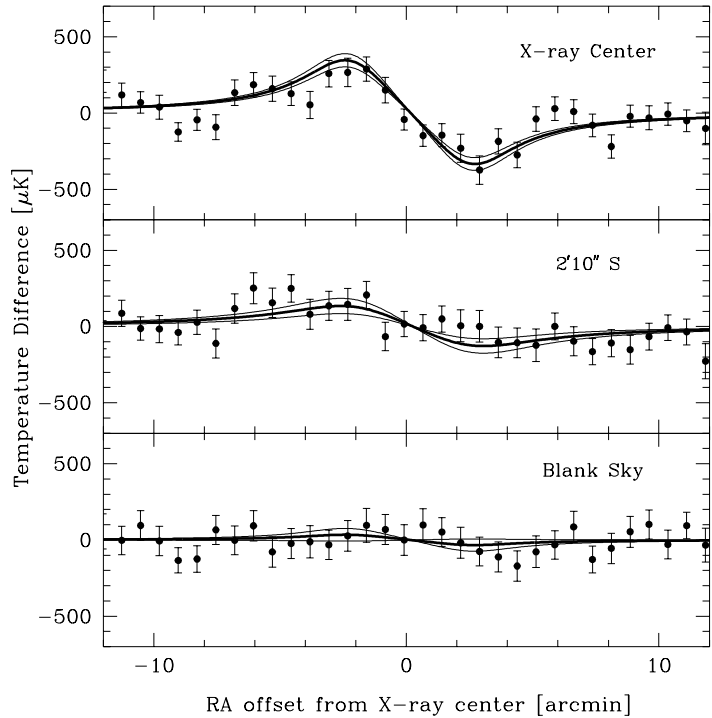}}
\caption{Coadded SuZIE drift scans across the X-ray center of Abell~2163,
$2'10''$ to the South, and on a patch of sky believed
to be free of sources. The line is the predicted SZE differential signal
calculated from the X-ray data assuming isothermal ICM gas, which has
been scaled to fit the 
data.\cite{holzapfel97a}}
\label{fig:SuZIE_a2163}
\end{figure}

All observations sensitive enough to observe the \sze\ are
differential.  The primary issue for single dish observations is how
to switch the beam on the sky without introducing systematics
comparable to the SZE.  This beam switching can be accomplished in
several ways, including but not limited to Dicke switching between
feeds and chopping mirrors which switch or sweep the beam on the sky.
With a single dish telescope, modulation of the beam sidelobes can
lead to an offset.  This offset can be removed if it remains stable
enough that it can be measured on a portion of the sky without the
cluster.  However, as a source is tracked over the course of an
evening, temperature variations of the optics and features on the
ground can cause the offset to change.  Therefore, it has become
common practice to observe leading and trailing fields with
the same position with respect to the ground as for the cluster observation.
In this
way, any constant or linear drift in offset can be removed at the
price of observing efficiency and sensitivity.  This technique has
been used successfully with the OVRO 5 meter telescope at 32 GHz to
produce reliable detections of the \sze\ in several
intermediate redshift clusters.\cite{herbig95,myers97} The SEST 30
meter and IRAM 15 meter telescopes have been used with bolometric
detectors at 140 GHZ and chopping mirrors to make significant
detections of the SZE in several
clusters.\cite{desert98,andreani96}  In the Sunyaev-Zel'dovich
Infrared Experiment (SuZIE) on the CSO submillimeter telescope,
pixels in the six element 142 GHz
bolometer array are electronically differenced by reading them out in
a differential bridge circuit.\cite{holzapfel97} Differencing in this
way makes the experiment insensitive to temperature and amplifier gain
fluctuations that produce 1/f noise.  This increased low frequency
stability allows SuZIE to observe in a drift scanning mode where the
telescope is fixed and the rotation of the earth moves the beams
across the sky.  Using this drift scanning technique, the SuZIE
experiment has produced high signal to noise strip maps of the SZE
emission in several clusters.\cite{holzapfel97b,mauskopf00} An example
of a SuZIE scan pattern and strip map are shown in
Fig.~\ref{fig:SuZIEscans} and Fig.~\ref{fig:SuZIE_a2163}, respectively.

Because of the high sensitivity of bolometric detectors at millimeter
wavelengths, single dish experiments are ideally suited for the
measurement of the SZE spectrum.  By observing at several
millimeter frequencies these instruments are able to separate the
thermal and kinetic SZE from atmospheric fluctuations and
sources of astrophysical confusion.  One of the first steps to
realizing this goal is the measurement of the SZE as an increment.  So
far, there have been only a few low signal to noise detections at a
frequency of approximately 270 GHz.  The main reason for the lack of
detection is the increased opacity of the atmosphere at higher
frequencies, and these measurements typically require the lowest
opacity and most stable atmosphere.  Holzapfel et al.\
(1997) \cite{holzapfel97b} report a detection of A2163 with the SuZIE
instrument at 270 GHz.  Observations by Andreani et al.\
(1996) \cite{andreani96} claim detections of the SZE increment in two
clusters, although observations of third cluster appear to be
contaminated by foreground sources or systematic 
errors.\cite{andreani99}

Single dish observations of the \sze\ are just beginning to reach
their potential and the future is promising.  The development of
large format millimeter wavelength bolometer arrays will increase the
mapping speed of current SZE experiments by orders of magnitude.  The
first of this new generation of instruments is the BOLOCAM 151 element
bolometer array \cite{mauskopf00b,glenn98} which has just completed a
engineering run at the CSO and will begin observations in May 2001.
BOLOCAM will observe in drift scanning mode and produce differences
between bolometer signals in software.  To the extent that atmospheric
fluctuations are common across the array, it will be possible to
realize the intrinsic sensitivity of the detectors.  Operating from
astronomical sites with stable atmospheres and low precipitable water
vapor, future large format bolometer arrays have the potential to
produce high signal to noise SZE images and search for distant SZE
clusters with unprecedented speed.

\subsection{Interferometric observations}
\label{sec:interferometer}

The stability and spatial filtering inherent to interferometry has
been exploited to make high quality images of the
SZE.\cite{jones93,grainge93,carlstrom96,grainge96,carlstrom98,patel00,joy01,grego99,grego00a,reese00,grego00b}
We first review the basic operating principles of an interferometer.

\begin{figure}[!t]
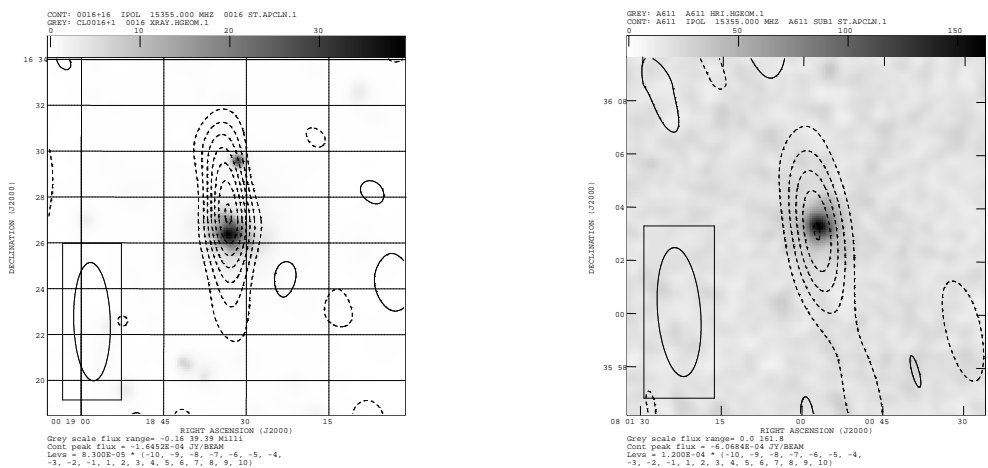

\begin{center}
\hbox to \columnwidth{
\hskip0.2in
\hbox to 3.0in{\epsfysize=2.50in\epsfbox{ryle_1.eps}}
\hfil
\hbox to 3.0in{\epsfysize=2.5in\epsfbox{ryle_2.eps}}
\hskip0.2in
}
\caption{ 
Ryle Telescope (RT) images (contours) superposed on ROSAT PSPC images
(greyscale) of the SZE effect in two clusters, CL~0016+16 at z=0.54 and Abell 611
at z=0.288. A611 has only 1/6 the X-ray luminosity of CL~0016+16. The ellipses
show the FWHM of the RT synthesized beam, which is strongly elongated in both
cases.\cite{jones}
\label{fig:ryle}}
\end{center}
\end{figure}

The `beam' of a two-element interferometer -- all arrays can be
thought of as a collection of $n(n-1)/2$ two-element interferometers
-- is essentially a cosine corrugation on the sky, analogous to a two
slit interference pattern and thus provides simultaneous differential
measurements.  The interferometer effectively multiplies
the sky brightness at the observing frequency by a cosine, integrates
the product and outputs the time average amplitude of the product.  In
practice the signal path is split and a 90 degree relative phase shift
is inserted into one path so that the output of the interferometer is
the complex Fourier transform of the sky
brightness at a spatial frequency given by $B/\lambda$, where $B$ is
the component of the vector connecting the two telescopes (the
baseline) oriented perpendicular to the source. Of course, a range of
baselines are actually being used at any one time due to the finite
size of the apertures of the individual array elements; this simply
reflects that the sky has been multiplied by the gain pattern (beam)
of the individual telescopes or, equivalently, that the Fourier
transform measured is the transform of the true sky brightness
convolved with the transform of the beam of an array
element.\footnote{the beam here is often referred to as the primary beam,
or gain envelope, of the interferometer and sets the field of view of
the interferometer at $\approx \lambda/D$ where $D$ is the diameter
of the telescope.}

The transformed beam is the auto-convolution of the aperture;
it is identically zero beyond the diameter of the telescopes expressed
in wavelengths. The interferometer is therefore only sensitive to
angular scales (spatial frequencies) near $B/\lambda$. It is
insensitive to gradients in the atmospheric emission or other large
scale emission features.  There are several other features which allow
an interferometer to achieve extremely low systematics. For example,
only signals which correlate between array elements will lead to
detected signal. For most interferometers, this means that the bulk of
the sky noise for each element will not lead to signal. Amplifier gain
instabilities for an interferometer will not lead to large offsets or
false detections, although if severe they may lead to somewhat noisy
signal amplitude. To remove the effects of offsets or drifts in the
electronics as well as the correlation of spurious (non-celestial)
sources of noise, the phase of the signal received at each telescope
is modulated before the correlator and then the proper demodulation is
applied to the output of the correlator.

\begin{figure}[!thb]
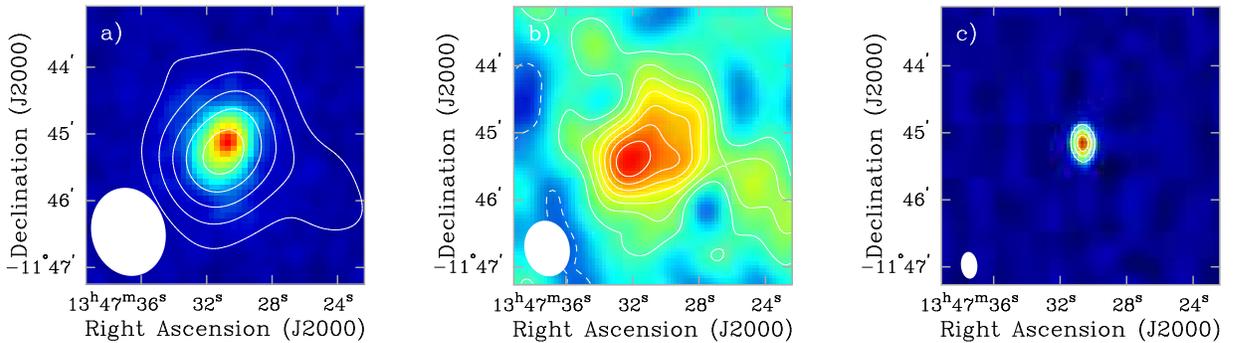

\begin{center}
\hbox to \columnwidth{
\hbox to 0.3\columnwidth{\epsfxsize=0.3\columnwidth
\epsfbox{r1347_szx_sqrt.ps}}
\hskip0.05\columnwidth
\hbox to 0.3\columnwidth{\epsfxsize=0.3\columnwidth
\epsfbox{r1347_highres.ps}}
\hskip0.05\columnwidth
\hbox to 0.3\columnwidth{\epsfxsize=0.3\columnwidth
\epsfbox{r1347_ptsrc.ps}}
}
\caption{
BIMA deconvolved interferometric images of the SZE of galaxy cluster
RXJ1347-11 emphasizing different spatial scales.  The FWHM ellipse of
the synthesized beam is shown in the lower left corner of each panel.
(a) Point source subtracted SZE image (contours) overlaid on 
 ROSAT X-ray image (false color).
The contours are multiples of 185$\mu$K ($\sim 2\sigma$), and negative contours are
shown as solid lines. A 1500 $\lambda$ half-power radius
Gaussian taper was applied resulting in a $63'' \times 80''$ synthesized beam.
The X-ray image is
HRI raw counts smoothed with a Gaussian with $\sigma = 6''$ and
contains roughly 4000 cluster counts.  (b) Higher resolution point
source subtracted SZE image
(both contours and false color).  A 3000 $\lambda$ half-power
radius Gaussian taper was applied resulting in a $40'' \times 50''$
synthesized beam. The contours are multiples of 175$\mu$K ($\sim 1\sigma$)
(c) Image of the point source made 
using projected baselines greater than 3000 $\lambda$.  This map
has a synthesized beam of $15'' \times 24''$ and a rms of
$\sim 275\ \mu$Jy beam$^{-1}$ ($1200\mu$K) sensitivity.  The contours are multiples of 10$\sigma$.}
\label{fig:rxj1347}
\end{center}
\end{figure}

Lastly, the spatial filtering of an interferometer allows the emission
from radio point sources to be separated from the \sze\ emission
(e.g., see Fig.~\ref{fig:rxj1347}). This is possible because at high
angular resolution ($\sim 10''$) the \sze\ contributes very little flux.
This allows one to use long baselines -- which give high angular
resolution -- to detect and monitor the flux of radio point sources
while using short baselines to measure the \sze.  Nearly simultaneous
monitoring of the point sources is important as they are often time
variable.  The signal from the point sources is then easily removed,
if they are not too strong, from the short baseline data which are
sensitive to the \sze. In practice, the fluxes of the point sources 
are fitted jointly with the SZE signal. 

For the reasons given above, interferometers offer an ideal way to
achieve high brightness sensitivity for extended low-surface
brightness sources, at least at radio wavelengths.  Most
interferometers, however, were not designed for imaging low-surface
brightness sources.  Interferometers are traditionally built to obtain
high angular resolution with large individual elements for maximum
sensitivity to small scale emission. Galaxy clusters, on the other
hand are large objects.
As a result, special purpose interferometric systems have been built
for imaging the SZE. All of them have taken advantage of low-noise
HEMT amplifiers \cite{pospieszalski95} to achieve high sensitivity.

The first interferometric detection \cite{jones93} of the SZE was
obtained with the Ryle
Telescope.\cite{jones93,grainge93,grainge96,grainge00} It is composed
of eight 13 m telescopes located in Cambridge, England operating at
15~GHz with East-West configurations.  Images of the SZE toward two
clusters obtained with the Ryle Telescope are shown in
Fig.~\ref{fig:ryle}; the East-West array configuration of the
telescope lead to North-South elongated synthesized beams.

%
%

\goodbreak
\begin{figure}[!t]
\centerline{\psfig{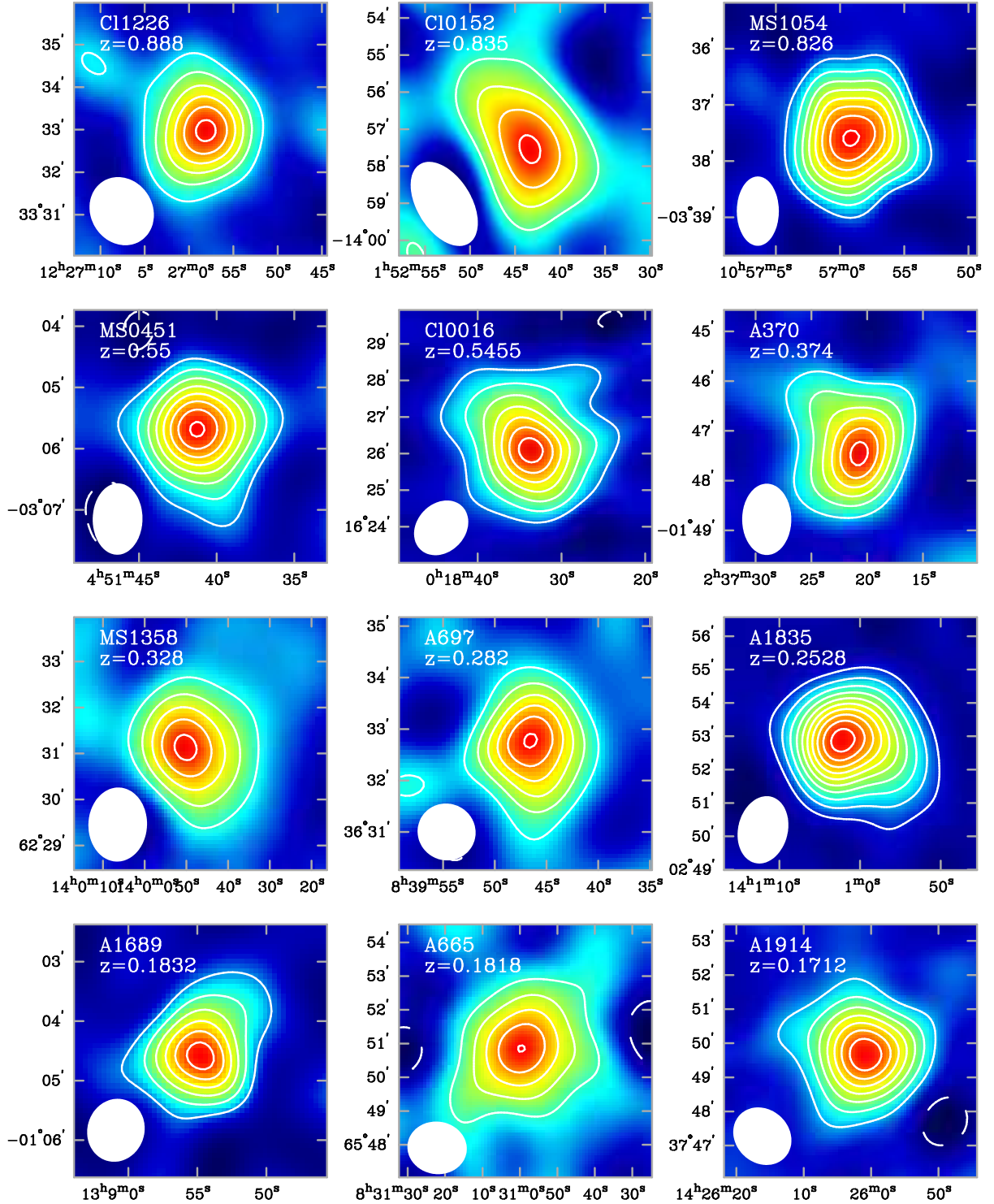}}
\caption{Deconvolved interferometric SZE images for a sample of galaxy
clusters over a large redshift range ($0.17 \leq z \leq 0.89$).  The
contours are multiples of 2$\sigma$ and negative contours are shown
as solid lines.  The FWHM ellipse of the synthesized beam is shown in
the lower left corner of each panel. The noise level $\sigma$ 
ranges from $20\mu$K to
$40\mu$K for the clusters shown.
}
\label{fig:szpanel12}
\vskip 24pt
\end{figure}
\goodbreak

The OVRO and BIMA SZE imaging project
\cite{carlstrom96,carlstrom99,grego99,patel00,grego00a,reese00,grego00b,joy01}
uses 30~GHz (1 cm) low noise receivers mounted on the OVRO
\footnote{An array of six 10.4 m telescopes located in the Owens
Valley, CA and operated by Caltech} or BIMA \footnote{An array of ten
6.1 m mm-wave telescopes located at Hat Creek, California and operated
by the Berkeley-Illinois-Maryland-Association} mm-wave arrays in
California. A subset of the 35 images obtained to date from the OVRO
and BIMA SZE imaging project is shown in Fig.~\ref{fig:szpanel12}. The
OVRO and BIMA arrays support two dimensional configurations of the
telescopes, including extremely short baselines, allowing good
synthesized beams for imaging the SZE of clusters at declinations
greater than $\sim -20$~degrees.

Fig.~\ref{fig:szpanel12} also clearly demonstrates the independence of
the \sze\ on redshift. The clusters have similar X-ray luminosities,
but span a factor of five in redshift. As can
be seen, the strength of the \sze\ for
each cluster is similar.

Point sources were removed from a large fraction of the SZE images
shown in Figs.~\ref{fig:ryle} and \ref{fig:szpanel12}. The
interferometers image the point sources simultaneously with the SZE by
using telescope configurations that include a large range of baseline
lengths. The point sources are identified in a straight forward manner
since their detected flux is independent of baseline length.  An
example of the effectiveness of the removal of bright point sources is
illustrated in Fig.~\ref{fig:rxj1347} where we show BIMA data of the
extremely X-ray luminous cluster RX~J1347-1145 at $z = 0.45$. As shown
in the right panel, a strong (10.7 mJy) point source is observed
toward the center of the cluster.  The SZE of this cluster has been
the target of single dish observations with the Nobeyama 45~m
telescope at 21~GHz, 43~GHz,\cite{komatsu99} and
150~GHz,\cite{komatsu00} with the DIABOLO bolometric system on the
IRAM 30~m at 140~GHz (2.1~mm) and 250~GHz (1.2~mm)
\cite{pointecouteau99} and with the SCUBA bolometric array on the JCMT
15~m telescope at 350~GHz.\cite{komatsu99} These observations
were not able to account for the contaminating
flux of the radio point source without adding considerable uncertainty
to the magnitude and morphology of the SZE.
The interferometric
observations presented here easily separated the
point source emission from the SZE. A recent high resolution
image made with the DIABOLO instrument on the IRAM telescope was
also able to distinguish the point source emission 
from the SZE.\cite{pointecouteau01}

\begin{figure}[t]
\epsfysize= 3 in
\centerline{\epsfbox{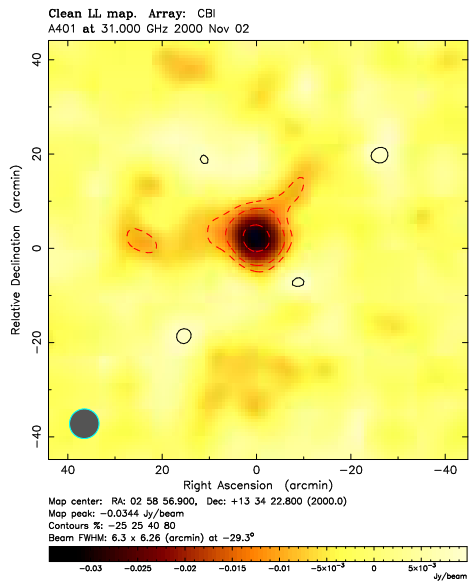}}
\caption{An image of the SZE toward Abell~401 obtained with the 
Cosmic Background Imager. The FHMW resolution is $6.3' \times 6.3'$.
The contour levels are at -25, 25, 40, and 80\% of the peak
level of $-34.2$ mJy per beam which corresponds to
a brightness temperature of 305$\mu$K. The noise level
is $20\mu$K.  
\cite{udomprasert00}}
\label{fig:a401_CBI}
\end{figure}

The Ryle, OVRO, and BIMA SZE observations are insensitive to the
angular scales required to image low redshift clusters, $z < 0.1$.
Recently, however, the Cosmic Background Imager (CBI) \cite{padin01}
has been used to image the SZE in a number of nearby clusters.  The
CBI is composed of thirteen 0.9 m telescopes mounted on a common
platform with baselines spanning 1 m to 6 m. Operating in ten 1 GHz
channels spanning 26 - 36 GHz, it is sensitive to angular scales
spanning 3$'$ to 45$'$. In Fig.~\ref{fig:a401_CBI} we show a CBI image
of the SZE toward Abell~401 at redshift 0.07.\cite{udomprasert00} As
for single dish observations, data from observations of leading and
trailing fields offset by 12.5 minutes in Right Ascension have been
subtracted from the CBI cluster image to remove contamination from
ground emission.

\section{Current Status of Cosmological Constraints from SZE measurements}
\label{sec:current_status}

Current SZE observations are of sufficient quality that already
significant constraints on cosmological parameters have been obtained.

\subsection{Hubble Constant}
\label{sec:hubble}

As outlined in section \ref{sec:cosmo-hubble}, by analyzing a
measurement of the SZE for a cluster in combination with its X-ray
properties, one can solve for the angular diameter distance to the
cluster. The Hubble constant is then easily computed with knowledge of
the cluster redshift.

\begin{figure}[t]
   \epsfysize = 3.0 in
   \centerline{\epsfbox{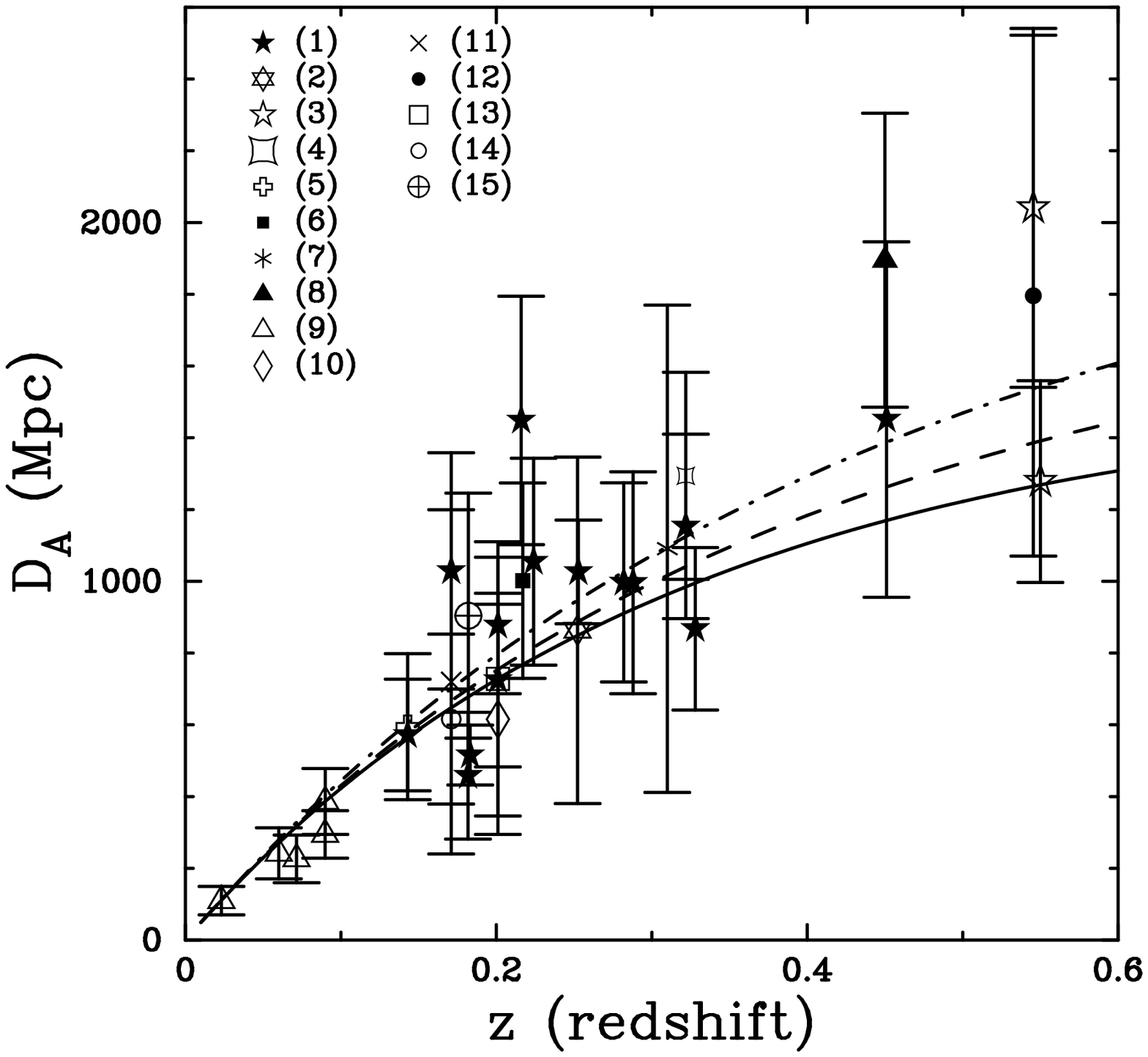}}
   \label{fig:da}
   \caption{SZE determined distances versus redshift. Also plotted
is the theoretical
   angular diameter distance relation for three different
   cosmologies, assuming $\Ho = 60$ \ksM.
References:
    (1)  \cite{reese00a},
    (2)  \cite{mauskopf00},
    (3)  \cite{reese00},
    (4)  \cite{patel00},
    (5)  \cite{grainge00},
    (6)  \cite{saunders00},
    (7)  \cite{andreani99},
    (8)  \cite{komatsu99},
    (9)  \cite{mason99,myers97,herbig95},
    (10) \cite{lamarre98},
    (11) \cite{tsuboi98},
    (12) \cite{hughes98b},
    (13) \cite{holzapfel97a},
    (14) \cite{birkinshaw94}, and
    (15) \cite{birkinshaw91}.}  
\end{figure}

The beauty of this technique for measuring the Hubble constant is that
it is completely independent of other techniques, and that it can be
used to measure distances at high redshifts.  While the method depends
only on well understood properties of fully ionized plasmas, there are
several sources of uncertainty in the derivation of the Hubble
constant for any particular cluster.  The largest uncertainty is 
the assumption that the cluster size along the line of
sight is comparable to its size in the plane of the sky. For this
reason it is desirable to use a large survey of clusters to determine
$H_o$.

The error budget in the determination of the Hubble constant is also
particularly sensitive to the electron temperature and measured SZE 
as it depends on the square of these quantities, i.e., 
\begin{equation}
H_o \propto \frac{S_X T_e^2}{(\Delta T_{SZE})^2}
\label{eq:Hpropto}
\end{equation}
where $S_X$ is the X-ray intensity.  Eq.~\ref{eq:Hpropto} also
illustrates the sensitivity of the SZE Hubble constant determination
to the {\it absolute calibration} of the X-ray and SZE observations.
Currently the best absolute calibration of SZE observations is $\sim
2.5$\% at 68\% C.L. based on observations of the brightness of the
planets, usually Mars and Jupiter.\cite{grego99,mason99}  This translates into a 5\%
systematic uncertainty shared by nearly all of the SZE measurements
made to date.  Efforts are now underway to reduce this uncertainty to
the 1\% level (2\% in $H_o$).  Uncertainty in the X-ray intensity
scale also adds another shared uncertainty as nearly all of the
reported SZE Hubble constant determinations use data from the ROSAT
satellite. The accuracy of the ROSAT intensity scale is debated, but a
reasonable estimate is believed to be $\sim 10$\%. It is hoped that
the calibration of the Chandra and XMM-Newton X-ray telescopes will greatly reduce
this uncertainty.

The current status of SZE determined distances is summarized in
Fig.~\ref{fig:da} along with the theoretical \Da\ relation for three
different cosmologies assuming $\Ho = 60$ \ksM.  It is clear that SZE
determined distances are beginning to trace the shape of the theoretical
angular diameter distance relation.
Excluding results from the BIMA and OVRO SZE project, there are
currently 16 SZE determined distances to 14 different clusters for
which reasonably high signal to noise SZE data and X-ray data exist.
We are currently analyzing the subset (17 clusters) of the 35 galaxy
clusters imaged with OVRO and BIMA SZE system for which good X-ray
data exists, roughly doubling the number of direct distances to galaxy
clusters (10 of which have no other SZE determined distance
estimates).  Preliminary results from 14 of these clusters can be
found in the article by Reese et~al. in this volume.\cite{reese00b}

A fit to all the SZE determined distances implies $\Ho = 63 \pm 3$
\ksM\ for an $\OmM=0.3$ and $\OmL=0.7$ cosmology, where the
uncertainty is the formal statistical uncertainty at 68\% confidence.
We caution that the statistical uncertainties are difficult to
interpret and the systematics are difficult to ascertain because many
of these distances share systematics.  Though the systematics are difficult to determine
precisely, they contribute roughly a 30\% uncertainty.
Many of the clusters are at high redshift where the
geometry of the universe affects the best fit Hubble constant.  These
33 distances imply a Hubble constant of $\Ho = 60$ \ksM\ for an open
$\OmM=0.3$ universe and $\Ho = 58$ \ksM\ for a flat $\OmM=1$
cosmology.  


As the quality of SZE and X-ray data improves, it will be possible to
employ ICM models more sophisticated than the spherical $\beta$-models
currently used.  Nevertheless, inspection of Fig.~\ref{fig:da} already
provides confidence that a survey of SZE distances consisting
of perhaps a few hundred clusters with redshifts extending to one and
beyond would allow the technique to be used to trace the expansion
history of the universe, providing a valuable independent check of the
type Ia supernova results.\cite{riess98,perlmutter99}

\subsection{Gas Masses, Baryonic Mass Fractions, and $\Omega_M$}
\label{sec:fg}

As outlined in Sec.~\ref{sec:cosmo_f_gas}, the ratio of the gas mass
of a cluster to its total gravitational mass provides a lower limit to its
total baryon fraction.  After compensating for the baryons contained
in the galaxies and those lost during cluster formation, the gas mass
fraction should reflect the universal baryon fraction
$\Omega_B/\Omega_M$.  Therefore a determination of the gas mass
fraction combined with constraints on $\Omega_B$ leads to an estimate
of $\Omega_M$. Here we use $\Omega_B h_{100}^2 = 0.019 \pm 0.002$ determined
from observations of D/H in Ly$\alpha$ clouds
\cite{burles98a,burles98b} along with recent reanalysis of big bang
nucleosynthesis predictions.\cite{nollett00,burles99} Such an estimate
of $\Omega_M$ is still strictly an upper bound, as we cannot rule out
the possibility of additional reservoirs of baryons in galaxy clusters
which have yet to be detected.

Two samples of SZE clusters have been analyzed and the results used to
place constraints on $\Omega_M$; a sample of four nearby clusters
\cite{myers97} and a sample of 18 distant
clusters.\cite{grego99,grego00b} Both analyses used a spherical
isothermal $\beta$-model for the ICM.  The nearby sample was observed
with the Owens Valley 5.5 m telescope at 32 GHz as part of a SZE study
of an X-ray flux limited sample.\cite{myers97} In this study, the
integrated SZE is used to normalize a model for the gas density from
published X-ray analyses, and this gas mass is compared to the
published total masses to determine the gas mass fraction within radii of 1-1.5
$h_{100}^{-1}$ Mpc. For three
clusters, A2142, A2256 and the Coma cluster, a gas mass
fraction of $(0.061\pm0.011) h_{100}^{-1}$ is found; for the cluster Abell 478, a gas mass
fraction of $(0.16\pm0.014)h_{100}^{-1}$ is reported.

The high redshift sample of 18 clusters ($0.14 < z < 0.83$) was
observed interferometrically at 30 GHz using the OVRO and BIMA SZE
imaging system.\cite{grego99,grego00b} In this study the model for the
gas density was determined directly by the SZE data; no X-ray imaging
data was used. X-ray derived emission weighted temperatures were used,
however.  The gas mass fractions were computed from the data at a
1$^\prime$ radius where they are best constrained by the observations.
Numerical simulations suggest, however, that the gas mass fraction at
$r_{500}$ (the radius at which the enclosed mean density of the cluster is 500 times
the critical density) should reflect the universal baryon 
fraction.\cite{evrard97,evrard96,david95} The derived gas mass fractions were
therefore extrapolated to $r_{500}$ using scaling relations from
cluster simulations.\cite{evrard97}

\begin{figure}[!t]
\epsfysize= 2.0 in
\centerline{\epsfbox{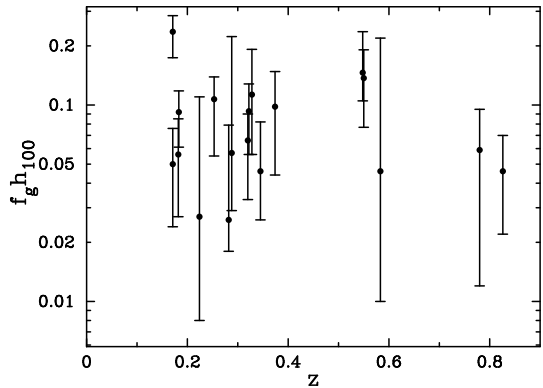}}
\caption{The gas mass fraction scaled to $r_{500}$ for galaxy clusters
derived from OVRO and BIMA Sunyaev-Zel'dovich effect data as a function of 
redshift assuming a 
$\Omega_M = 0.3,\ \Omega_\Lambda=0.7$ cosmology.\cite{grego99,grego00b} The mean gas mass fraction 
is $f_g = 0.081 ^{+0.009} _{-0.011} \ h_{100}^{-1}$.}
\label{fig:fg}
\end{figure}

The resulting gas mass fractions assuming a $\Omega_M = 0.3,\
\Omega_\Lambda=0.7$ cosmology for each cluster are shown as a function
of redshift in Fig.~\ref{fig:fg}.  The uncertainty in the electron
temperatures contribute the largest component to the error budget. The
resulting mean gas mass fractions are 
$f_g = 0.081 ^{+0.009} _{-0.011}\ h_{100}^{-1}$ for $\Omega_M = 0.3,\ \Omega_\Lambda=0.7$, 
$f_g = 0.074^{+0.008} _{-0.009} \ h_{100}^{-1}$ for $\Omega_M = 0.3,\ \Omega_\Lambda=0.0$ and 
$f_g = 0.068 ^{+0.009} _{-0.008} \ h_{100}^{-1}$ for $\Omega_M = 1.0,\ \Omega_\Lambda=0.0$.

Gas mass fractions derived from X-ray images for a large, homogeneous,
nearby sample of clusters are presented in 
Mohr et al.\ (1999).\cite{mohr99}  For a
subsample of 28 clusters with $T_e > 5$ keV, they find the mean gas
mass fraction within $r_{500}$ to be $(0.0749\pm
0.0021)h_{100}^{-3/2}$ at 90\% confidence.  The gas mass fractions
derived from SZE measurements depend differently on the cosmology
assumed than those derived from X-ray images, and this should be noted
when comparing the results.  Qualitatively, the comparison does not
suggest any large systematic offsets.  This is significant, because a
large clumping factor, $C \equiv <n_e^2>^{1/2}/<n_e>$, has been
suggested as an explanation for the high gas mass fractions in
clusters.\cite{white93,evrard97} A cluster with clumping factor $C$
would only require 1/$C$ as much gas mass to produce the observed
emission, and the SZE and X-ray gas mass fraction measurements would
differ by a factor of $\sim C$.

\begin{figure}[!h]
\epsfysize= 3 in
\centerline{\epsfbox{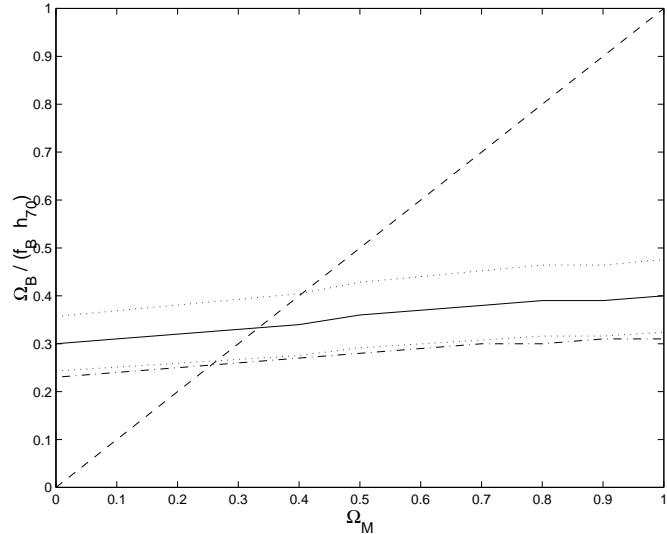}}
\caption{Limits on $\Omega_M$ from SZE measured cluster gas
fractions.\cite{grego00b} Upper limit on the total matter density,
$\Omega_M \le \Omega_B/(f_B h_{70})$ (full line) and its associated
68\% confidence region (dotted lines), as a function of cosmology with
$\Omega_\Lambda \equiv 1 - \Omega_M$.  The intercept between the upper
dotted line and the dashed line $\Omega_M = \Omega_B/(f_B h_{70})$
gives the upper limit to $\Omega_M$ at 68\% confidence.  The
dot-dashed line shows the total matter density when the baryon
fraction includes an estimate of the contribution from baryons in
galaxies and those lost during cluster formation.  The intercept of
the dot-dashed line and the dashed line gives the best estimate of
$\Omega_M \sim 0.25$ assuming a flat universe with $h =0.7$.}
\label{fig:fg_omega}
\end{figure}

The measured gas mass fractions can be used to determine $\Omega_M$ in
a self-consistent manner. Fig.~\ref{fig:fg_omega} shows the value of
$\Omega_M$ implied by the measured gas mass fractions assuming a flat
universe ($\Omega_\Lambda \equiv 1- \Omega_M$) and $h=0.7$ to
calculate the angular diameter distance and $r_{500}$ scaling factor.
The upper limit to $\Omega_M$ and its associated 68\% confidence
interval is shown as a function of $\Omega_M$.  The measured gas mass
fractions are consistent with a flat universe and $h=0.7$ when
$\Omega_M$ is less than 0.40, at 68\% confidence.  For the
measurements to be consistent with $\Omega_M = 1.0$ in a flat
universe, the Hubble constant must be very low, $h$ less than $\sim$
0.30.

For a more realistic estimate, an attempt is made to account for the
baryon contribution from galaxies and those lost during cluster
formation.  The mass of the cluster galaxies is assumed to be a fixed 
fraction of the
cluster gas mass, with the fraction fixed at the value observed in the Coma
cluster, $M_{gas} = M_g(1+0.20 h_{100}^{3/2})$.\cite{white93} The gas in the
cluster will be more extended than the dark matter and the baryon
fraction at $r_{500}$ will be a modest underestimate of the true
baryon fraction $f_g(r_{500}) = 0.85\times f_b({\rm
universal})$.\cite{evrard97} These assumptions lead to $f_B=(f_g(1 +
0.2h_{100}^{3/2})/ 0.85)$ Using this to scale the gas mass fractions derived from
the high z cluster sample and assuming $h=0.7$ and a flat cosmology,
leads to the constraints illustrated in Fig.~\ref{fig:fg_omega}.  The
best estimate of $\Omega_M$ is $\sim$0.25.

\subsection{Cluster Peculiar Velocities}
\label{sec:peculiar}

The kinetic SZE signal from the peculiar velocity of galaxy clusters
is very small ($\leq 0.1$ mK) and difficult to detect.  Disentangling
the thermal and kinetic SZE's requires multi-frequency SZE
observations.  To date there have been only a few recent attempts to
measure the effect.

\begin{figure}[!t]
\begin{center}
\hbox to \columnwidth{
\hskip0.2in
\hbox to 2.5in{\epsfxsize=2.5in\epsfbox{BIMA_a2163_szxray.ps}}
\hfil
\hbox to 3.0in{\epsfxsize=3.0in\epsfbox{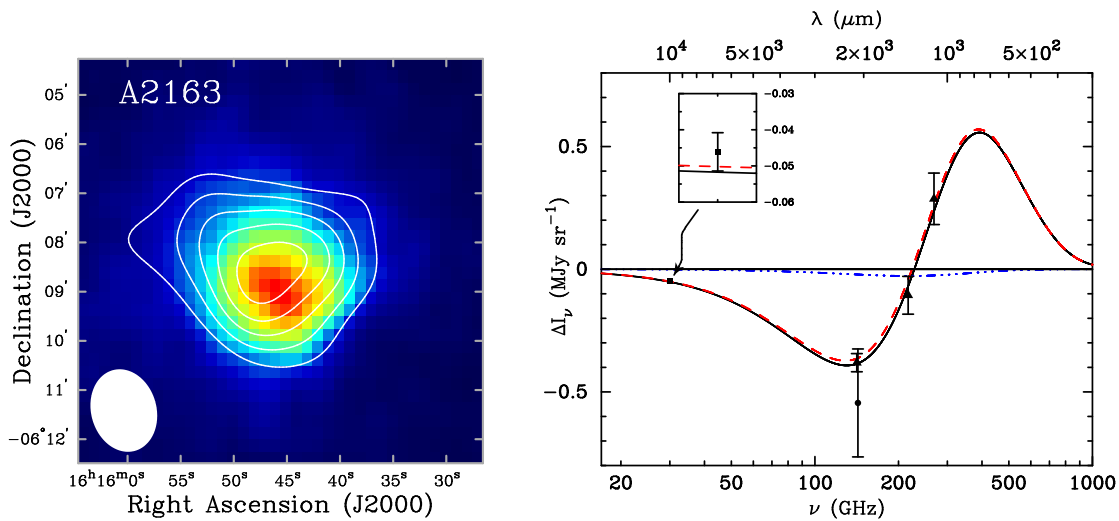}}
\hskip0.2in
}
\caption{ Left panel: An image of the SZE effect toward Abell~2163
obtained with the BIMA interferometer overlaid on a ROSAT image of the
X-ray emission. Right panel: the SZE spectrum of Abell 2163 using data
from BIMA at 28.5~GHz,\cite{laroque01} DIABOLO at 140~GHz
\cite{desert98} (filled circle) and SuZIE at 140~GHz, 218~GHz and
270~GHz \cite{holzapfel97b} (filled triangles).  The best fit thermal
and kinetic SZE spectra are shown by the dot-dashed line and the
dashed lines, respectively, with the spectra of the combined effect
shown by the solid line. The limits on the Compton $y$-parameter and
the peculiar velocity are $y_0=3.65\pm0.40\times 10^{-4}$ and
$v_p=415^{+920}_{-765} \mbox{ km
s}^{-1}$.\cite{holzapfel97b,laroque01} }
\label{fig:a2163_spec}
\end{center}
\end{figure}

The first interesting limits on the peculiar velocity of a galaxy
cluster were reported in Holzapfel et al.\ (1997).\cite{holzapfel97b}
They used SuZIE to observe Abell~2163 ($z = 0.201$) and Abell~1689 ($z
= 0.181$) at 140~GHz (2.1 mm), 218~GHz (1.4 mm) and 270~GHz (1.1
mm). These observations include and bracket the null in the thermal
SZE spectrum.  Using a $\beta$ model, with the shape parameters
($\theta_c$, $\beta$) from X-ray data, they find $V_p = +490 ^{+1370}
_{-880}$ \kms\ for Abell 2163 and $V_p = +170 ^{+815}_{-630}$ \kms\
for Abell 1689, where the uncertainties are at 68\% confidence and
include both statistical and systematic uncertainties.  These results
are limited by the sensitivity of these SZE observations, which are
limited by differential atmospheric emission. The SuZIE data for
Abell~2163 have been reanalyzed with the addition of higher frequency
measurements which are sensitive to emission from Galactic dust in the
direction of the cluster.\cite{lamarre98} More recently LaRoque et
al.\ (2000) \cite{laroque01} also reanalyzed all of the available data
for Abell~2163, including a new measurement obtained with the OVRO and
BIMA SZE imaging system at 30 GHz (1 cm). As shown in
Fig.~\ref{fig:a2163_spec}, they find the data is well fitted by
parameters remarkably similar to the original values from Holzapfel et
al.\ with insignificant contamination by Galactic dust emission when
the SuZIE observing scheme is taken into account.

\section{Summary}

We have reviewed the large amount of progress made in 
detecting and imaging the SZE over the last several years. The
SZE has lived up to its initial promise of providing an
independent estimate of the Hubble constant. It has also
been used to estimate the matter density of the universe.
The full potential of the SZE as a cosmological tool, however,
is largely unrealized; the SZE is ideally suited for
conducting large, deep, surveys for clusters. As discussed
in Sec.~\ref{sec:cosmo-survey}, SZE surveys are particularly
powerful since the cluster detection threshold
for such a survey is the cluster mass and is essentially
independent of redshift.

SZE surveys will provide a direct view of structure
formation in the high-redshift universe; if clusters exist at 
redshifts much higher than
currently predicted they will be found by the SZE survey, but missed in
even the deepest X-ray observations planned.

A deep SZE
survey can expect to find roughly one cluster per square degree with
$z>1.5$. The large sample of high redshift clusters found in 
such a survey can be followed up with X-ray observations and
then used to refine  SZE $H_o$ estimates and even constrain
the deceleration parameter, providing a valuable check 
of the type Ia supernova results.\cite{riess98,perlmutter99}

Most importantly, using the data from SZE surveys
to determine the evolution of the number density and mass distribution
of clusters will lead to tight constraints on $\Omega_M$, $\Omega_\Lambda$,
and even the equation of state for the dark energy.\cite{haiman00}

Lastly, surveys with fairly high angular resolution ($\sim 30'')$ will
be sensitive to the evolution of the intracluster medium,\cite{holder01}
perhaps seeing the effects of
feedback from galaxy formation out to $z\sim 2$.

While the present instruments and
techniques 
have led to remarkable progress
(see Secs.~\ref{sec:single_dish} and \ref{sec:interferometer}),
they are much
too slow to conduct the large, deep, SZE surveys we
envision. In some ways, the current instruments are
prototypes of the instruments needed to 
conduct large SZE surveys. 
Due to the success of these instruments, we now have the
knowledge to design powerful SZE survey instruments. Both
dedicated interferometric arrays and single dish telescopes equipped
with large bolometric arrays are likely to conduct deep surveys
over tens of square degrees in the near future.

SZE survey machines that are currently being built or awaiting
funding include interferometric arrays
(SZA by the OVRO -- BIMA SZE imaging group, AMI by
the Ryle SZE imaging group, AMiBA in Taiwan, 
and upgrades to CBI and DASI) and 
bolometric arrays (ACBAR on the Viper telescope at
the South pole, BOLOCAM on the CSO telescope on Mauna Kea, 
and a large format bolometer array on a 
$\sim 10$m telescope located at the South Pole (SPST)).
All of these instruments will be capable of surveying
tens of square degrees, if not much more, per year. The
interferometers are likely to provide the highest
resolution and deepest images, while the bolometric arrays are
likely to provide the largest sky coverage. Both
will be needed to fully exploit the power of
the SZE as a cosmological tool.

\section*{Acknowledgment}

This work is supported in part by NASA LTSA grant NAG5-7986.  
We thank Mike Joffre, Amber Miller, and Daisuke Nagai for
helpful comments and discussions. JEC
acknowledges generous support from the David and Lucile Packard
Foundation and the James S. McDonnell Foundation. We thank
the conference organizers for an enjoyable and stimulating
conference.

\section*{References}


\end{document}